\documentclass[sigconf,natbib=false,authorversion=true]{acmart}

\AtBeginDocument{%
  }

\usepackage{bm}
\usepackage{tikz}
\usetikzlibrary{positioning}
\usepackage{multirow}
\usepackage{rotating}
\usepackage[caption=false]{subfig}
\usepackage[noend]{algpseudocode}
\usepackage{algorithm}
\usepackage{hyphenat}

\RequirePackage[
  datamodel=acmdatamodel,
  style=acmnumeric,
  ]{biblatex}

\copyrightyear{2024}
\acmYear{2024}
\setcopyright{acmlicensed}
\acmConference[GECCO '24]{Genetic and Evolutionary Computation Conference}{July 14--18, 2024}{Melbourne, VIC, Australia}
\acmBooktitle{Genetic and Evolutionary Computation Conference (GECCO '24), July 14--18, 2024, Melbourne, VIC, Australia}
\acmDOI{10.1145/3638529.3653988}
\acmISBN{979-8-4007-0494-9/24/07}

\begin{document}

\title[Learning from Evolution]{Learning from Evolution: Improving Collective Decision\hyp{}Making Mechanisms using Insights from Evolutionary Robotics}

\author{Tanja Katharina Kaiser}
\email{tanja.kaiser@utn.de}
\orcid{0000-0002-1700-5508}
\affiliation{%
  \institution{University of Technology Nuremberg}
  \city{Nuremberg}
  \country{Germany}
}

\begin{abstract}
Collective decision\hyp{}making enables multi\hyp{}robot systems 
to act autonomously in real\hyp{}world environments. 
Existing collective decision\hyp{}making mechanisms suffer from the so\hyp{}called speed versus accuracy trade\hyp{}off or rely on high complexity, e.g., by including global communication. 
Recent work has shown that more efficient collective decision\hyp{}making mechanisms based on artificial neural networks can be generated using methods from evolutionary computation. 
A major drawback of these decision\hyp{}making neural networks is their limited interpretability.  
Analyzing evolved decision\hyp{}making mechanisms can help us improve the efficiency of hand\hyp{}coded decision\hyp{}making mechanisms while maintaining a higher interpretability. 
In this paper, we analyze evolved collective decision\hyp{}making mechanisms in detail and hand\hyp{}code two new decision\hyp{}making mechanisms based on the insights gained. 
In benchmark experiments, we show that the newly implemented collective decision\hyp{}making mechanisms are more efficient than the state\hyp{}of\hyp{}the\hyp{}art collective decision\hyp{}making mechanisms voter model and majority rule. 
\end{abstract}


\maketitle

\section{Introduction}

\begin{figure}[t]
    \centering
    \includegraphics[width=0.6\linewidth]{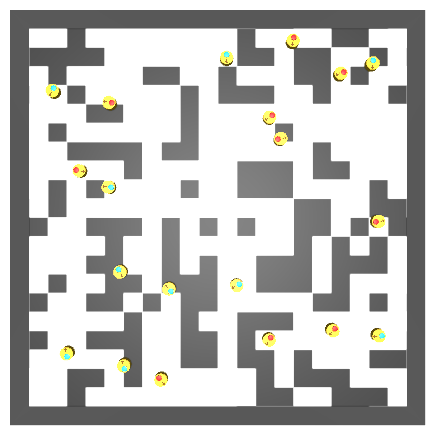}
    \caption{Collective perception scenario in the BeeGround simulator.  
    The arena is bounded by walls and the environmental features are represented by the black and white pattern on the arena floor. Robots indicate their current opinion via an LED on top (red stands for \textit{Black}, blue for \textit{White}).}
    \label{fig:arena}
\end{figure}

Collective decision\hyp{}making allows multi\hyp{}robot systems to cooperate in complex environments by collectively making choices. 
Often, the best\hyp{}of\hyp{}\(n\) options~\cite{Valentini2017} must be chosen, where each robot has only incomplete information about all available options and can only communicate with its local neighborhood. 
Still, collective decisions should be fast and accurate. 
Speed and accuracy are considered to be conflicting goals in collective decision\hyp{}making (i.e., speed versus accuracy trade\hyp{}off)~\cite{valentini2016collective}. 
Faster decision\hyp{}making leads to a loss of decision accuracy and higher accuracy slows down decision\hyp{}making. 
This is illustrated by two state\hyp{}of\hyp{}the\hyp{}art collective decision\hyp{}making mechanisms: voter model and majority rule. 
Each robot adopts the opinion of a random neighbor in its local neighborhood when using the voter model, and the majority opinion of its neighbors when using the majority rule. 
The voter model is accurate but slow; the majority rule is less accurate but fast. 

Researchers have proposed a variety of new decision\hyp{}making mechanisms to improve the efficiency of collective decision\hyp{}making.
Often, the collective perception scenario introduced by Valentini et al.~\cite{Valentini2016} is used as an example problem setting. 
In this scenario,  robots must collectively decide which of two environmental features is the more frequent, i.e., choose the best of two options (best\hyp{}of\hyp{}\(n\) problem with \(n = 2\)). 
In their paper introducing the collective perception scenario, Valentini et al.~\cite{Valentini2016} compare their previously proposed self\hyp{}organizing decision\hyp{}making strategies \textit{Direct Modulation of Majority\hyp{}based Decisions} (DMMD)~\cite{Valentini2015} and \textit{Direct Modulation of Voter\hyp{}based Decisions} (DMVD)~\cite{Valentini2014}. 
In both approaches, positive feedback is modulated by longer dissemination phases for high\hyp{}quality opinions, and a specific decision\hyp{}making mechanism (i.e., the majority rule and the voter model, respectively) is applied. 
The authors compare their approaches to the \textit{Direct Comparison} (DC) of option quality strategy where robots disseminate the quality estimate of their current opinion. 
A robot will switch to the opinion of a random neighbor only if the quality estimate of the neighbor's opinion is higher than the robot's own quality estimate. 
Thus, the DC strategy relies on more information from a robot's neighbors than the DMMD and DMVD strategies resulting in communication overhead.
Compared to the self\hyp{}organizing strategies DMMD and DMVD, the DC strategy is more efficient for simple settings in small multi-robot systems but faces more severe scalability issues regarding consensus time with increasing problem difficulty and multi-robot system size.
Bartashevich and Mostaghim~\cite{Paline2019} couple the direct modulation of positive feedback with an Ising\hyp{}based collective decision\hyp{}making mechanism. 
Robots can switch between different decision\hyp{}making mechanisms (i.e., parameterizations of the Ising model) based on their current internal preference for the overall collective decision. 
The authors found that the Ising model mostly performs similarly to DMMD. 
Shan and Mostaghim~\cite{Shan2021} propose \textit{Distributed Bayesian Belief Sharing} as a new collective decision\hyp{}making mechanism that outperforms DMVD and DC at the cost of increased communication complexity.

Other researchers automatically generate new decision\hyp{}making strategies using evolutionary computation methods~\cite{eiben_2003_introduction}. 
Almansoori et al.~\cite{almansoori2021evolution, Almansoori2022} evolve neural networks as decision\hyp{}making strategies using a task\hyp{}specific fitness function. 
In a comparative study, the authors show that the evolved decision\hyp{}making strategies outperform the voter model. 
In our previous work~\cite{kaiser2023}, we evolve decision\hyp{}making mechanisms to be used in the direct modulation of positive feedback decision\hyp{}making strategy of Valentini et al.~\cite{Valentini2016} using a task\hyp{}specific, a task\hyp{}independent, and a hybrid fitness function.
We show that the decision\hyp{}making mechanisms evolved with the task\hyp{}specific fitness function and the hybrid fitness function outperform voter model and majority rule. 
In both works, the decision\hyp{}making approaches lack interpretability due to their representation as artificial neural networks. 
In contrast, hand\hyp{}coded decision\hyp{}making mechanisms such as the voter model and the majority rule are easily explainable and transparent. 
However, analyzing the evolved decision\hyp{}making mechanisms allows us to draw inspiration for new, improved hand\hyp{}coded decision\hyp{}making mechanisms. 
In this paper, we 

\begin{enumerate}
    \item analyze evolved decision\hyp{}making mechanisms in detail and investigate the impact of the input features on the decision output,  
    \item develop two new hand\hyp{}coded decision\hyp{}making mechanisms based on the knowledge gained, 
    \item and show the effectiveness of the newly implemented decision\hyp{}making mechanisms in comparison to the voter model and the majority rule. 
\end{enumerate}

All figures and additional material are available on Zenodo~\cite{zenodo_2024}.

\section{Method}

\subsection{Collective Perception Scenario}

\subsubsection{Arena}
We simulate the collective perception scenario
in the BeeGround simulator~\cite{Lim2021} using the same  arena and robot parameters as in the original collective perception experiments by Valentini et al.~\cite{Valentini2016}.  
The arena has a size of~\(2~\text{m} \times 2~\text{m}\) and is bounded by walls. 
The two environmental features are represented by black and white tiles of \(10~\text{cm} \times 10~\text{cm}\) each, see Fig.~\ref{fig:arena}. 
Task difficulty~\(\rho^*\)  is defined as 
\begin{equation}
	\rho^* =  \min \left( \frac{\rho_\text{white}}{\rho_\text{black}}, \frac{\rho_\text{black}}{\rho_\text{white}} \right) \, , 
\end{equation}

where \(\rho_\text{white}\) and  \(\rho_\text{black}\) are the percentages of white and black tiles in the environment, respectively. 
A problem difficulty of~\(1\) means that half of the tiles are white and half of the tiles are black.  
We vary task difficulty~\(\rho^*\) by using different ratios of black and white tiles.

\subsubsection{Robots}
Each robot has a differential drive with a maximum speed of~\(10~\frac{\text{cm}}{\text{s}}\), an LED on top indicating its current opinion,  
and five horizontal proximity sensors to the front with a range of~\(10\)~cm. 
A binary ground sensor indicates whether the arena surface below is black~(\(0\)) or white~(\(1\)). 
Robots can broadcast their current opinion in the range of~\(70~\text{cm}\) (i.e., local communication) and store the opinions of up to four unique neighbors in a message queue. 
The information stored in the message queue~\(Q\) is aggregated into two virtual sensors. 
We normalize both sensors to reduce the dependency on the queue size. 
Sensor~\(l(t)\) gives the number of received messages normalized by the maximum length of the message queue. 
It is defined as 

\begin{equation}
    l(t) = \frac{|Q(t)|}{4} \, , 
\end{equation}

where \(|Q(t)|\) is the length of the message queue \(Q\) at time step~\(t\).
Sensor~\(w(t)\) gives the percentage of neighbors with opinion \textit{White}~(\(1\)) in the message queue (excluding the robot's own opinion). 
It is defined as 

\begin{equation}
    w(t) = \frac{|\{q \in Q(t):q = 1\}|}{|Q(t)|}\, . 
\end{equation}

We randomly distribute \(N=20\)~robots in the arena at the beginning of each run. 
Half of the robots are initialized with opinion~\textit{White}, the other half with opinion~\textit{Black}.

\subsubsection{Robot Motion and Decision\hyp{}Making Routines}

\begin{figure}[t]
    \centering
        \subfloat[decision\hyp{}making strategy \label{fig:psfm_decision}]{\resizebox{0.37\linewidth}{!}{
		 \tikzstyle{arrow} = [thick,->,>=stealth]
	   \begin{tikzpicture}[node distance=2cm]
		\node (exploration) [rectangle, rounded corners, minimum width=2.5cm, minimum height=0.6cm, text centered, draw=black, align=center] {exploration};
		\node (start) [above=0.75cm of exploration] { };
		\node (dissemination) [rectangle, rounded corners, minimum width=2.5cm, minimum height=0.6cm, text centered, draw=black, below=2.5cm of exploration, align=center] {dissemination};
		\node (decision) [rectangle, minimum width=2.5cm, minimum height=0.6cm, text centered, draw=black, above=0.55cm of dissemination, align=center] {decision\hyp{}making\\mechanism};
		\draw [arrow] (start) -- node[right,align=left]{start}  (exploration);
		\draw [arrow] (exploration) to [bend right=80, left, align=right] node{after\\$t_n^\text{exp}$} (dissemination);
		\draw [arrow] (dissemination) to [right,align=left] node{after $t_n^\text{dis}$} (decision);
		\draw [arrow,dotted] (decision) to [right,align=left] node{after \\ decision} (exploration);
	\end{tikzpicture}}}
   \hspace{3mm} 
    \subfloat[random walk motion routine \label{fig:fsm_motion}]{\resizebox{0.58\linewidth}{!}{
		\tikzstyle{arrow} = [thick,->,>=stealth]
			\begin{tikzpicture}[node distance=2cm]
				\node (motion) [rectangle, rounded corners, draw=black, minimum width=2.25cm, minimum height=0.6cm,text centered] {straight motion};
				\node (start) [above=0.5cm of motion] { };
				\node (avoidance) [rectangle, rounded corners, minimum width=2.25cm, minimum height=0.6cm,text centered, draw=black, below=of motion,align=center] {obstacle\\avoidance};
				\node (unstuck) [rectangle, rounded corners, minimum width=2.25cm, minimum height=0.6cm,text centered, draw=black, right=of avoidance] {unstuck};
				\node (rotation) [rectangle, rounded corners, minimum width=2.25cm, minimum height=0.6cm,text centered, draw=black, right=of motion] {rotation};
				
				\draw [arrow] (start) -- (motion);
				\draw [arrow] (avoidance) to [above] node{$\zeta_n < 0$} (unstuck);
				\draw [arrow] (motion) to [bend left=25, below right, align=left] node{obstacle\\detected} (avoidance);
				\draw [arrow] (avoidance) to [bend left=25, left, align=right] node{turned\\by $\beta_n$} (motion);
				\draw [arrow] (motion) to [bend right=25, above,align=center] node{after $t^{\textrm{str}}_n$} (rotation); 
				\draw [arrow] (rotation) to [bend right=25, below,align=center] node{after $t^{\textrm{rot}}_n$} (motion);
				\draw [arrow] (unstuck) to node[below,align=left, xshift=2.3cm]{no obstacle\\detected} (motion);
			\end{tikzpicture}}}
    \caption{Probabilistic finite state machine for decision\hyp{}making~(a) and finite state machine for robot motion~(b) based on Valentini et al.~\cite{Valentini2016}. Time periods~\(t^{exp}_n\), \(t^{dis}_n\), \(t^{rot}_n\), \(t^{str}_n\) and angle~\(\beta_n\) are randomly sampled. Buffer values~\(\zeta_n < 0\) indicate that the robot is potentially stuck between obstacles.} 
    \label{fig:fsms}
\end{figure}
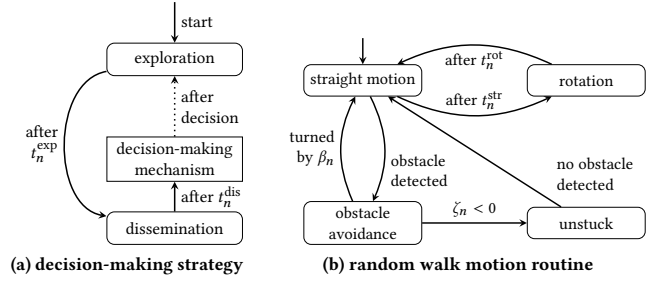

We closely follow the setup of Valentini et al.~\cite{Valentini2016} for the basic behavioral routines of the robots. 
On each robot, we concurrently execute a probabilistic finite state machine (PFSM) for decision\hyp{}making, which implements the decision\hyp{}making strategy and a random walk motion routine, see Fig.~\ref{fig:fsms}.

The decision\hyp{}making PFSM implements the direct modulation of positive feedback, see Fig.~\ref{fig:psfm_decision}. 
Each robot first explores its environment by sampling the local ground color for a time period $t_n^\textrm{exp}$ drawn from an exponential distribution with a mean of~$10~\textrm{s}$ to determine a quality estimate ~$\hat{\rho}_n$ for its current opinion.  
The longer a robot perceives the color matching its current opinion during the current exploration phase, the higher the quality estimate. 
In the subsequent dissemination phase, a robot broadcasts its current opinion to its local neighborhood at $1~\textrm{Hz}$ for a time period~$t_n^\textrm{send}$ sampled from an exponential distribution with a mean of $T_\textrm{send}\hat{\rho}_n$. 
In our experiments, design parameter $T_\textrm{send}$ is set to $10~\textrm{s}$. 
As a result, higher\hyp{}quality opinions lead to longer dissemination phases, resulting in a modulation of positive feedback.  
After sending its opinion, a robot receives the opinions of its neighbors for $t_n^\textrm{receive} = 3~\textrm{s}$. 
After $t_n^\textrm{dis} = t_n^\textrm{send} + t_n^\textrm{receive}$, the robot applies a decision\hyp{}making mechanism to update its current opinion. 
Different decision\hyp{}making mechanisms (i.e., voter model, majority rule, evolved decision\hyp{}making mechanisms, and the two new hand\hyp{}coded decision\hyp{}making mechanisms here) can be executed, allowing comparisons using the same overall decision\hyp{}making strategy. 
Finally, the robot returns to the exploration state. 

Simultaneously with the decision\hyp{}making PFSM, each robot performs a random motion behavior, alternating between straight motion and rotation on the spot in a random direction, see Fig.~\ref{fig:fsm_motion}. 
The random duration of straight motion is sampled from an exponential distribution with a mean of~$40~\textrm{s}$, and the random duration of rotation is sampled from a uniform distribution with bounds $[0~\textrm{s}, 4.5~\textrm{s}]$. 
A robot will rotate by $\beta_n = 180^\circ + x_\textrm{rand}$ where $x_\textrm{rand}$ is sampled from a uniform distribution with bounds $[-25^\circ, 25^\circ]$ if it detects an obstacle (i.e., a wall or another robot) while moving straight.  
The unstuck state is used to prevent thrashing and is enabled when a robot has recently turned most of the time, indicating that it may be stuck between obstacles. 
A buffer value of~$\zeta_n < 0$ means that a robot has recently been rotating mostly.  
The buffer value decreases when a robot is rotating, either in the rotation or obstacle avoidance state. 
The value increases up to a maximum of $7.5~\textrm{s}$ when the robot is moving straight. 
When the unstuck state is enabled, the robot will rotate in a random direction until no obstacles are detected.

\subsection{Benchmarks}
\label{sec:benchmarks}

For a fair comparison, we will benchmark the different decision\hyp{}making mechanisms using the same initial settings (i.e., initial robot poses and opinions, ground patterns of black and white tiles). 
We evaluate each collective decision\hyp{}making mechanism for \(400~\text{s}\) in \(1\,000\) different settings each for the more frequent feature~\textit{White} and the more frequent feature \textit{Black} per problem difficulty \(\rho^* \in \{0.25, 0.52, 0.67, 0.82 \}\). 
The ground patterns of black and white tiles and the initial robot opinions for \textit{White}\hyp{}dominant environments and for \textit{Black}\hyp{}dominant environments are mirrored. 
This way we can investigate whether the decision\hyp{}making mechanisms work equally well for both features.

\subsection{Evaluation Metrics}

To evaluate the efficiency of the collective decision\hyp{}making mechanisms,  we use two common metrics: mean consensus time~$\overline{T}_N$ to measure decision speed and exit probability~$E_N$ to measure decision accuracy. 
Consensus time~$T_N$ is the time it takes a multi-robot system to reach a first consensus (regardless of whether it is right or wrong) and mean consensus time~$\overline{T}_N$ is the average consensus time over all runs that lead to a consensus. 
Exit probability~$E_N$ is the percentage of runs in which \(100~\%\) of the robots successfully reached a consensus for the more frequent feature.

\subsection{Evolved Decision\hyp{}Making Mechanisms}
\label{sec:evolution}

\tikzset{%
	every neuron/.style={
		circle,
		draw,
		minimum size=0.7cm
	},
	neuron missing/.style={
		draw=none, 
		scale=2,
		text height=0.333cm,
		execute at begin node=\color{black}$\vdots$
	},
}

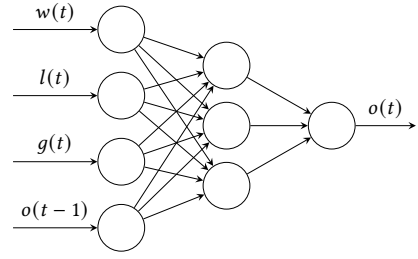
\begin{figure}[t]
		\resizebox {0.65\linewidth} {!} {
			\begin{tikzpicture}[x=0.9cm, y=0.9cm, >=stealth,]
				\foreach \m/\l [count=\y] in {1,2,3,4}
				\node [every neuron/.try, neuron \m/.try] (input-\m) at (0,2 -\y*1.1) {};
				\foreach \m [count=\y] in {1,2,3}
				\node [every neuron/.try, neuron \m/.try ] (hidden-\m) at (1.75,1.3-\y) {};
				\foreach \m [count=\y] in {1}
				\node [every neuron/.try, neuron \m/.try ] (output-\m) at (3.5,0.3-\y) {};
				\draw [<-] (input-1) -- ++(-1.8,0)
				node [above, midway] {$w(t)$};
				\draw [<-] (input-2) -- ++(-1.8,0)
				node [above, midway] {$l(t)$};
				\draw [<-] (input-3) -- ++(-1.8,0)
				node [above, midway] {$g(t)$};
				\draw [<-] (input-4) -- ++(-1.8,0)
				node [above, midway] {$o(t-1)$};
				\draw [->] (output-1) -- ++(1.4,0)
				node [above, midway] {$o(t)$};
				\foreach \i in {1,2,3,4}
				\foreach \j in {1,2,3}
				\draw [->] (input-\i) -- (hidden-\j);
				\foreach \i in {1,2,3}
				\foreach \j in {1}
				\draw [->] (hidden-\i) -- (output-\j);
		\end{tikzpicture} }
\caption{Topology of the evolved decision\hyp{}making mechanisms. Inputs are the percentage of neighbors with opinion \textit{White}~\(w(t)\), the normalized length of the message queue of neighbor opinions~\(l(t)\), the ground sensor value~\(g(t)\) at current time step~\(t\), and the robot's previous opinion~\(o(t-1)\). The ANN outputs the robot's current opinion~\(o(t)\). 
\label{fig:ANN} }
\end{figure}

We evolve the weights of three\hyp{}layer feedforward artificial neural networks (ANN), see Fig.~\ref{fig:ANN}, as decision\hyp{}making mechanisms using a simple evolutionary algorithm. 
Each ANN receives four inputs: 
(i)~the percentage of neighbors with opinion~\textit{White}~\(w(t)\), 
(ii)~the normalized queue length~\(l(t)\), 
(iii)~the current ground sensor value~\(g(t)\), 
and (iv)~the robot's previous opinion~\(o(t-1)\). 
The ANN outputs the robot's current opinion~\(o(t)\). 
As in our previous work~\cite{kaiser2023}, our fitness function~$F$ rewards higher percentages of robots with the correct opinion in the last time step~\(T-1\) of the evaluation. 
Fitness~$F$ is given by 
\begin{equation}
    F = \frac{1}{N}  \sum^{N-1}_{n=0}  f_n \, , 
   \label{equ:TS}
\end{equation}

with 

\begin{equation}
    f_n = \begin{cases}
        1 & \text{if } o_n(T-1)=1   \text{ and dominant feature} = \text{\textit{White}}\\
        & \text{or } o_n (T-1) = 0 \text{ and dominant feature} = \text{\textit{Black}}\\ 
        0 & \text{otherwise} 
            \end{cases} \,    , 
\end{equation}
size of the multi-robot system~\(N\), and opinion~\(o_n(T-1)\) of robot~\(n\) in the last time step~\(T-1\) of the run. 
Opinion~\(o_n\) is \(0\) for \textit{Black} and \(1\) for \textit{White}. 
An initial population of \(50\)~genomes is randomly generated that are evaluated on homogeneous multi-robot systems, i.e., each robot executes a copy of the same decision-making ANN.
Each genome is evaluated for \(200\)~s in three different environments with random black\hyp{}and\hyp{}white patterns of problem difficulty~$\rho^* = 0.25$ and in environments with the inverse of those patterns to avoid bias toward one feature (i.e., six evaluations in total). 
We restrict ourselves to problem difficulty~$\rho^* = 0.25$ since our previous work~\cite{kaiser2023} has shown that decision\hyp{}making mechanisms evolved in easier settings scale better with problem difficulty. 
The overall fitness of a genome is set to the minimum fitness observed in these six evaluations. 
We run evolution for \(600\)~generations, use fitness proportionate parent selection, age\hyp{}based survivor selection, elitism of~\(1\), a mutation rate of \(0.2\), and no crossover. 
In total, we do ten independent evolutionary runs.
For each evolutionary run, we select the best\hyp{}evolved individual of the last generation for a more detailed analysis.

\subsection{Analysis}
\label{sec:evaluation}

First, we analyze the performance (i.e., mean time to consensus and exit probability) of the best-evolved individual of each run relative to the voter model and the majority rule in our benchmark settings (see Sec.~\ref{sec:benchmarks}) to identify differences between the ten independent evolutionary runs. 
We then generate a global explanation of our evolved decision\hyp{}making mechanisms using the SHAP (SHapley Additive exPlanations) framework~\cite{shap}. 
We compute the mean feature importance measures using the shap Python package and all possible network input combinations to determine the impact of each input feature on the decision output.
Since our sensor inputs specifying the percentage of neighbors with opinion \textit{White} \(w(t)\) and the normalized length of the message queue~\(l(t)\) are highly correlated (i.e., possible values $w$ depend on the queue length~$l$), we group both input features in our analysis, as is done for correlated variables in other work~\cite{NOHARA2022106584}. 
In addition, we analyze the distribution of the actual values of the input features during an evaluation. 
This allows us to determine whether some input features have (mostly) fixed values. 
We use the insights gained to hand\hyp{}code two new decision\hyp{}making mechanisms (see Sec.~\ref{sec:new_cdm}), which offer higher interpretability than the evolved decision\hyp{}making mechanisms. 
Finally, we compare the two new hand\hyp{}coded decision\hyp{}making mechanisms to the voter model, the majority rule, and the evolved decision\hyp{}making mechanisms in the benchmark settings. 
For the overall comparison, we quantify the performance of the ten evolved decision\hyp{}making mechanisms using their median values for the mean consensus time~$\overline{T}_N$ and the exit probability~$E_N$. 
We test for statistical significance using the Mann-Whitney U test. 

\section{Results}

\subsection{Evolved Decision\hyp{}Making Mechanisms}

\subsubsection{Overall Performance}

We find a median best fitness of~\(1.0\) in the last generation of the evolutionary runs when rewarding higher percentages of robots with the correct opinion in the last time step of an evaluation (see Eq.~\ref{equ:TS}). 
The best\hyp{}evolved individuals of all \(10\)~evolutionary runs reach the correct consensus in all of their six evaluations, i.e., an exit probability~\(E_N\) of~\(100~\%\) is achieved. 
The mean consensus time~\(\overline{T}_N\) is within~\([51.2~\text{s}, 71.0~\text{s}]\) with a median of~\(54.1~\text{s}\). 
We assume that slower decision-making mechanisms will be removed by the evolutionary process, since not all robots may have the correct opinion at the end of all six evaluations for these mechanisms, leading to an implicit preference for faster decision-making mechanisms.  
Overall, evolution successfully optimizes the artificial neural networks as collective decision\hyp{}making mechanisms for problem difficulty~\(\rho^* = 0.25\).

\subsubsection{Efficiency of the Best\hyp{}Evolved Individuals}
\label{sec:efficiency}

\begin{figure*}[t]
    \centering
    \subfloat[$\overline{T}_N$ for \(\rho^* = 0.67\) \label{fig:TN_pR_067}]{\includegraphics[width=0.45\linewidth]{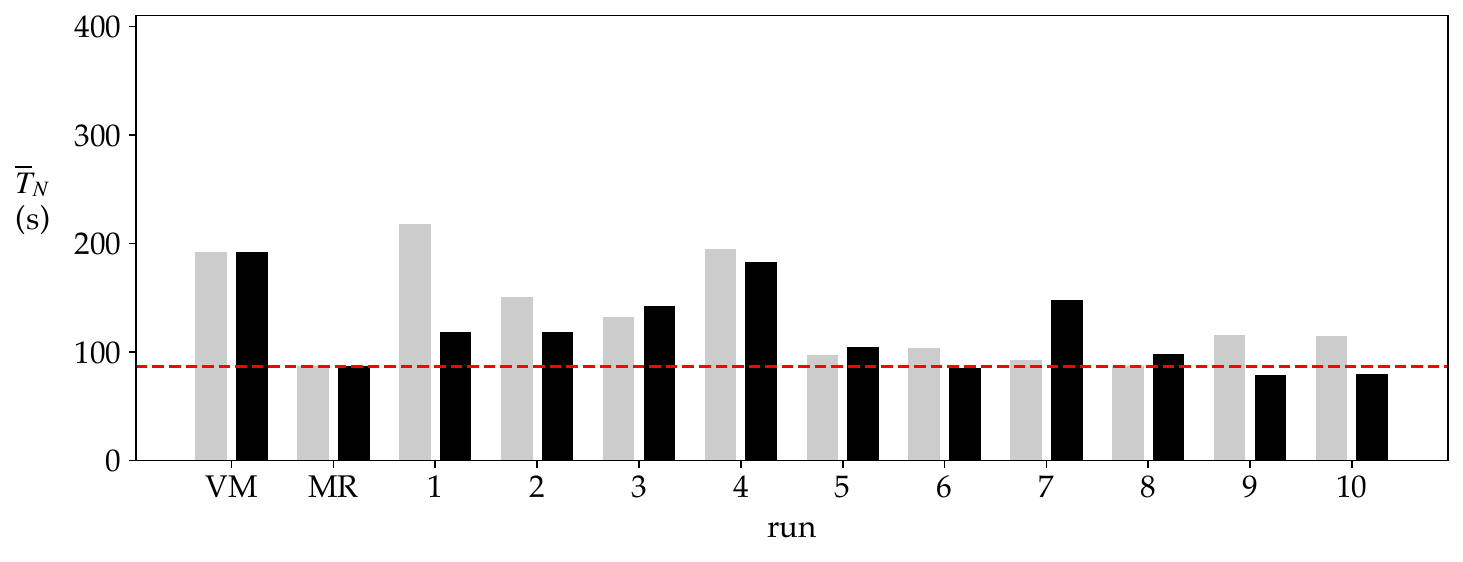}}
    \subfloat[$E_N$ for \(\rho^* = 0.67\) \label{fig:EN_pR_067}]{\includegraphics[width=0.45\linewidth]{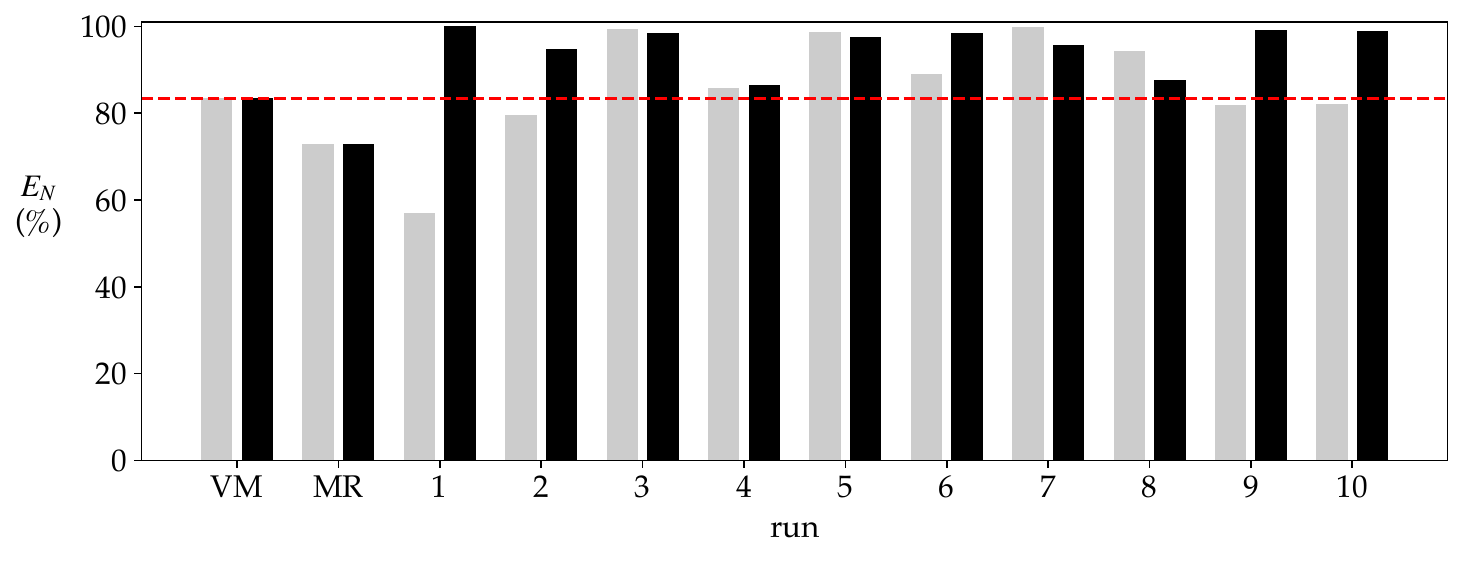}} \\
    \subfloat[$\overline{T}_N$ for \(\rho^* = 0.82\) \label{fig:TN_pR_082}]{\includegraphics[width=0.45\linewidth]{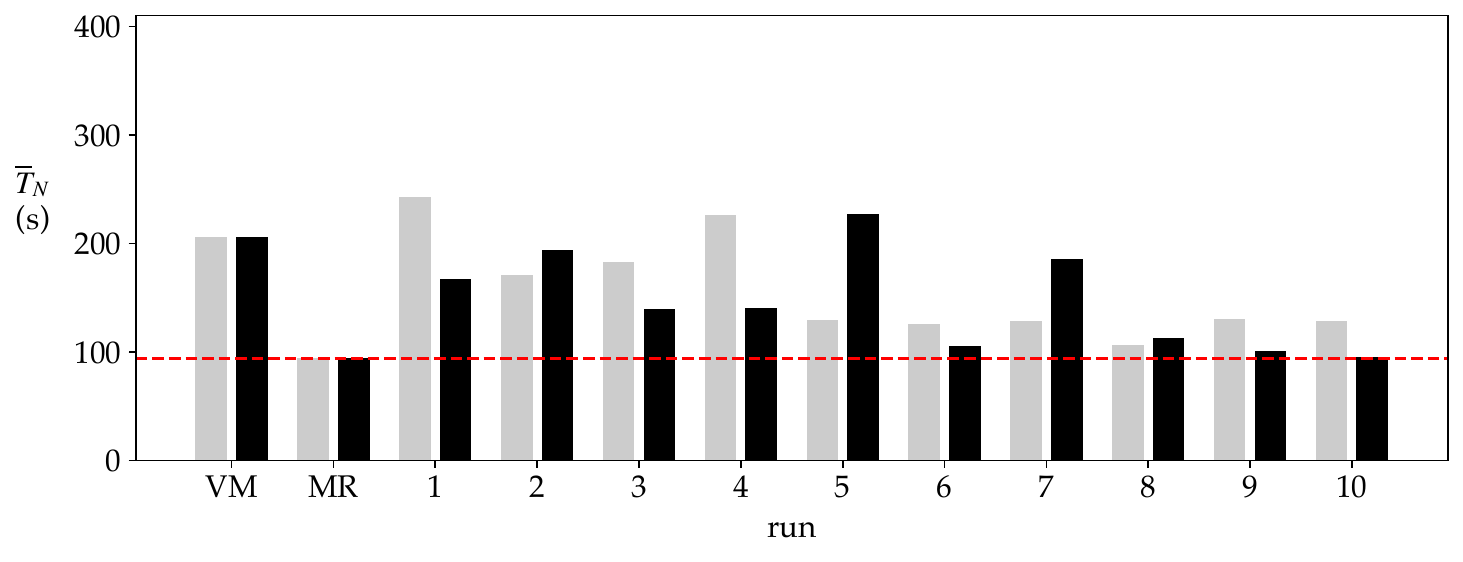}} 
    \subfloat[$E_N$ for \(\rho^* = 0.82\) \label{fig:EN_pR_082}]{\includegraphics[width=0.45\linewidth]{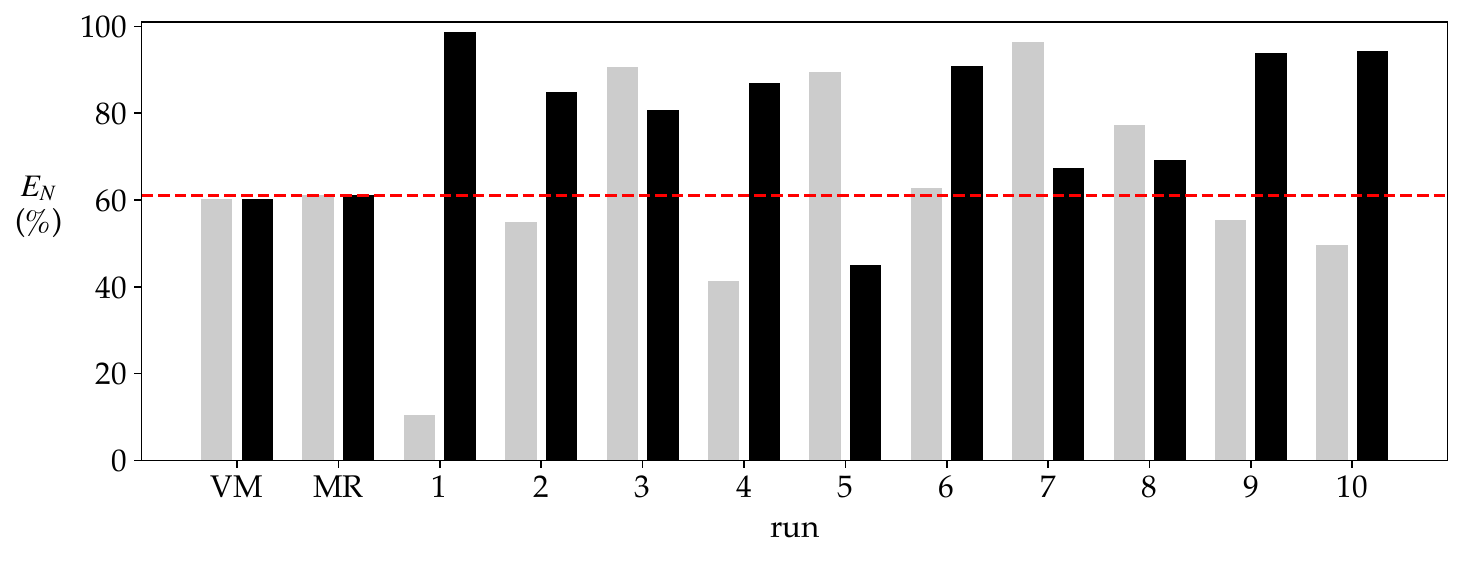}}
    \caption{Mean consensus time~$\overline{T}_N$ and exit probability~$E_N$ for each of the ten best\hyp{}evolved individuals in problem difficulties~\(\rho^* \in \{0.67, 0.82\} \).
    We split up the data between \textit{White}\hyp{}dominant (gray bars, left) and \textit{Black}\hyp{}dominant (black bars, right) environments. 
    Red dashed lines indicate the best performance of our baselines voter model and majority rule, i.e., the speed (i.e., lowest \(\overline{T}_N\)) of the majority rule and the accuracy (i.e., highest \(E_N\)) of the voter model.
    Plots for \(\rho^* \in \{0.25, 0.52\} \) are available on Zenodo~\cite{zenodo_2024}.}
    \label{fig:E_T_per_run}
\end{figure*}

In our benchmark settings (see Sec.~\ref{sec:benchmarks}), we compare the best\hyp{}evolved decision\hyp{}making mechanisms of the ten evolutionary runs against each other as well as against the voter model and the majority rule as a baseline, see Fig.~\ref{fig:E_T_per_run}. 
For problem difficulty~\(\rho^* = 0.25\) (i.e., the easiest setting), all evolved decision\hyp{}making mechanisms achieve \(100\)~\% decision accuracy for both \textit{White}\hyp{}dominant and \textit{Black}\hyp{}dominant environments.
Thus, they are as accurate as the voter model and slightly outperform the majority rule (\(E_N = 96.7~\% \)). 
Nine out of ten individuals are faster than the majority rule, and all ten individuals are faster than the voter model (p\(< 0.05\)). 
On average, the evolved decision\hyp{}making mechanisms are thus more efficient than the voter model and the majority rule for the problem difficulty used during the optimization process (see Sec.~\ref{sec:evolution}). 
With increasing problem difficulty, the number of best\hyp{}evolved individuals that outperform the voter model and the majority rule in decision accuracy decreases for both  \textit{White}\hyp{}dominant and \textit{Black}\hyp{}dominant environments. 
We find seven out of ten best\hyp{}evolved individuals for problem difficulty~\(\rho^* = 0.52\), six out of ten runs for problem difficulty~\(\rho^* = 0.67\) (see Fig.~\ref{fig:EN_pR_067}), and four out of ten runs for problem difficulty~\(\rho^* = 0.82\) (see Fig.~\ref{fig:EN_pR_082}) that outperform voter model and majority rule in both \textit{White}\hyp{}dominant and \textit{Black}\hyp{}dominant environments. 
Similarly, the number of best\hyp{}evolved individuals that outperform the majority rule and the voter model in terms of decision speed decreases with increasing problem difficulty. 
For problem difficulty~\(\rho^* = 0.52\), two out of the ten best\hyp{}evolved individuals outperform the majority rule (p\(< 0.05\)). 
For problem difficulties~\(\rho^* = 0.67\) and~\(\rho^* = 0.82\), none of the ten best\hyp{}evolved individuals outperform the majority rule in decision speed (p\(< 0.05\)). 
However, eight of the ten best\hyp{}evolved individuals in problem difficulty~\(\rho^* = 0.67\) (see Fig.~\ref{fig:TN_pR_067}) and seven of the ten best\hyp{}evolved individuals in problem difficulty~\(\rho^* = 0.82\) (see Fig.~\ref{fig:TN_pR_082}) outperform the voter model (p\(< 0.05\)). 

The voter model and the majority rule perform equally well for \textit{White}\hyp{}dominant and \textit{Black}\hyp{}dominant environments. 
For most of the evolved decision\hyp{}making mechanisms, we find increasing differences in decision speed and accuracy depending on whether \textit{White} or \textit{Black} is the most frequent feature as problem difficulty increases. 
The solutions are randomly biased for one of the two features potentially caused by ANN weights that favor one feature when the feature ratio is low.  
Especially for the higher problem difficulties~\(\rho^* = \{0.67, 0.82\}\), the differences in efficiency can be severe (p\(< 0.05\)). 
It should be noted that the decision\hyp{}making mechanisms were evolved in problem difficulty~\(\rho^*=0.25\) where no differences in decision accuracy (p\(> 0.05\)) and only in six out of ten runs significant differences in decision speed (p\(< 0.05\)) can be found depending on whether \textit{White} or \textit{Black} is the most frequent feature. 
Thus, not all evolved decision\hyp{}making mechanisms perform similarly well in different problem difficulties when rewarding higher percentages of robots with the correct opinion in the last time step.
A refinement of the fitness function and the evolution of collective decision\hyp{}making mechanisms in different problem difficulties during an evolutionary run could lead to improvements in this respect.
When drawing insights from the evolved collective decision\hyp{}making strategies for our new hand\hyp{}coded decision\hyp{}making mechanisms, we should weigh insights from the best\hyp{}evolved individuals that perform similarly for \textit{White}\hyp{}dominant environments and \textit{Black}\hyp{}dominant environments in all problem difficulties higher. 
We find that the best\hyp{}evolved individuals of evolutionary runs~3 and~8 perform best across problem difficulties in terms of similar exit probabilities~\(E_N\) and mean consensus times~\(\overline{T}_N\) for both 
\textit{White}\hyp{}dominant and \textit{Black}\hyp{}dominant environments, see Fig.~\ref{fig:E_T_per_run}. 
The best\hyp{}evolved individual of run~3 has higher exit probabilities~\(E_N\) than the best\hyp{}evolved individual of run~8 while the best\hyp{}evolved individual of run 8 leads to lower mean consensus times~\(\overline{T}_N\). 
This is consistent with the well\hyp{}known speed versus accuracy trade\hyp{}off, although some of the evolved decision\hyp{}making mechanisms may outperform voter model and majority rule in speed and accuracy. 
Next, we will analyze the impact of the input features on the decision output.
We will weigh the insights gained from the best\hyp{}evolved individuals of runs~3 and~8 higher but to get an overall impression of the input features favored by evolution, we will analyze and base our conclusions on all evolved decision\hyp{}making mechanisms.

\subsubsection{Impact of Input Features}
\label{sec:impact}

\begin{figure}[t]
    \centering
    \subfloat[run 2 \label{fig:shap2}]{\includegraphics[width=0.9\linewidth]{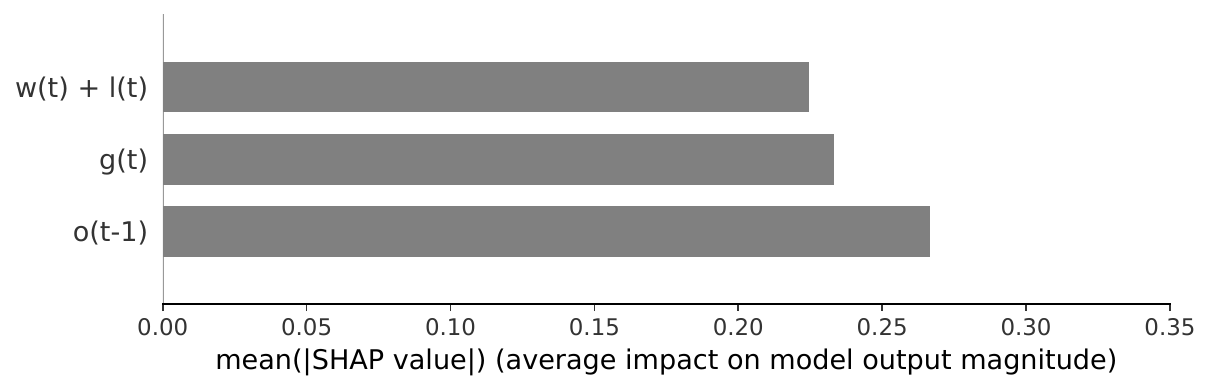}} \\
    \subfloat[run 3 \label{fig:shap3}]{\includegraphics[width=0.9\linewidth]{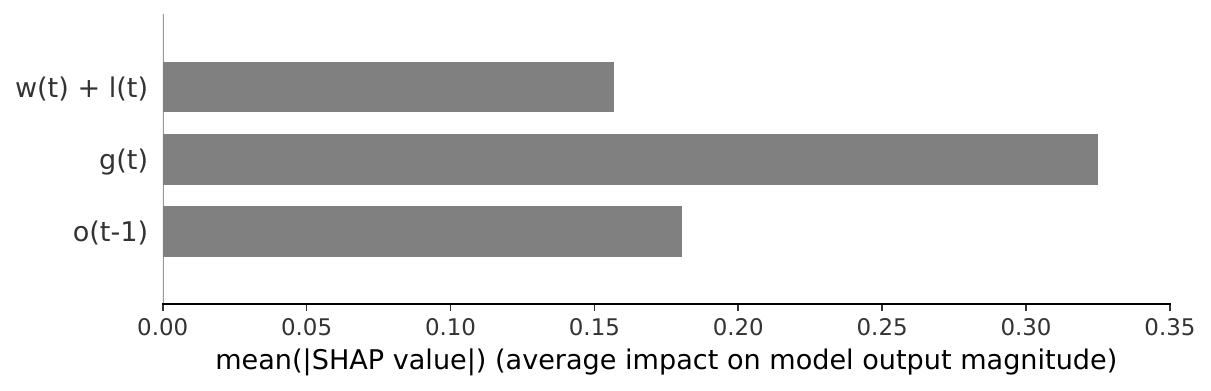}} \\
    \subfloat[run 6 \label{fig:shap6}]{\includegraphics[width=0.9\linewidth]{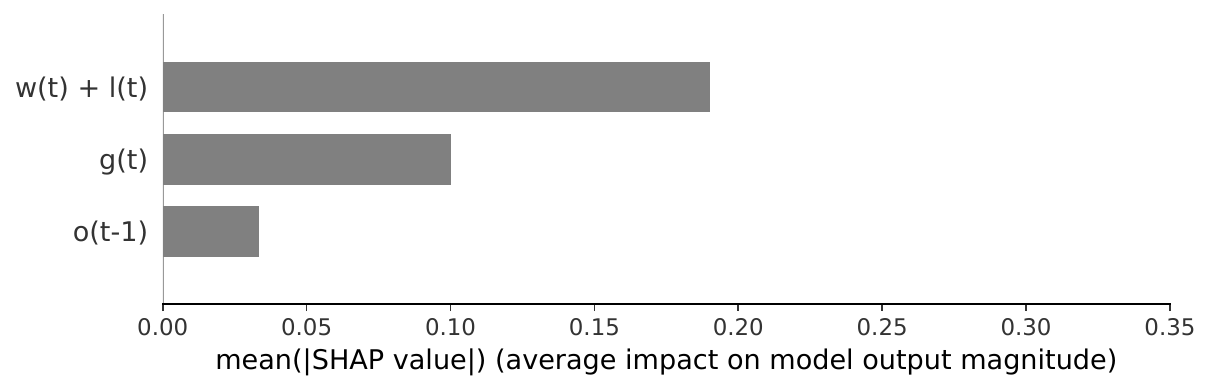}} \\
    \subfloat[run 8 \label{fig:shap8}]{\includegraphics[width=0.9\linewidth]{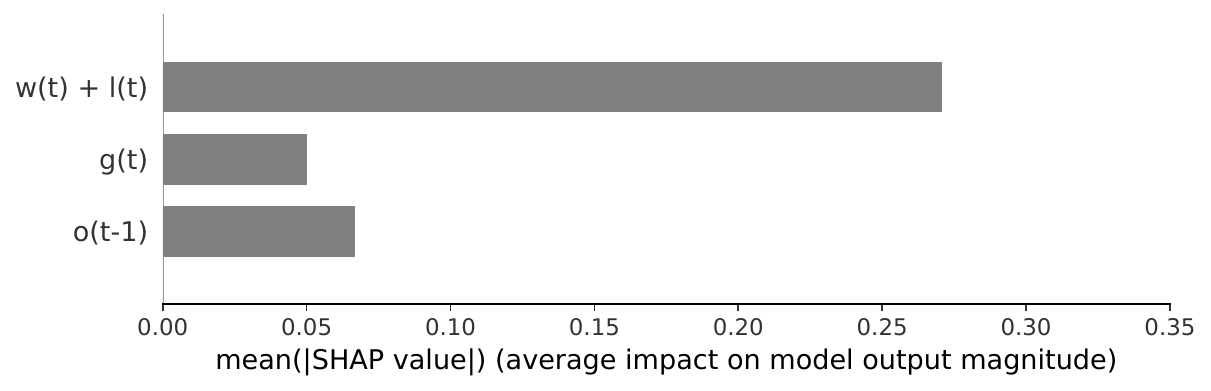}}
    \caption{Mean SHAP values for four representative runs where \(w(t)\)~is the percentage of neighbors with opinion \textit{White}, \(l(t)\)~the normalized length of the message queue of neighbor opinions, \(g(t)\)~the ground sensor value, and \(o(t-1)\)~the robot's previous opinion. \(w(t)\) and \(l(t)\) are grouped as their values are highly correlated. Plots for all other runs and beeswarm plots are available on Zenodo~\cite{zenodo_2024}.} 
    \label{fig:SHAP}
\end{figure}

We study the impact of the input features on the decision output by calculating the SHAP values for each of the ten best\hyp{}evolved collective decision\hyp{}making mechanisms. 
Overall, we find three different approaches for collective decision\hyp{}making in the evolved decision\hyp{}making mechanisms. 
All have similar efficiency. 

The first approach is used by the majority of best\hyp{}evolved individuals (i.e., seven out of ten) and depends most strongly on the opinions of a robot's neighbors, i.e., on the percentage of neighbors with opinion \textit{White}~\(w(t)\) and the normalized queue length~\(l(t)\). 
In six of these runs, the ground sensor value~\(g(t)\) has the second highest impact (see Fig.~\ref{fig:shap6}), and in one run the robot's previous opinion~\(o(t-1)\) ranks second (see Fig.~\ref{fig:shap8}).  
The influence of the ground sensor~\(g(t)\) and the robot's previous opinion~\(o(t-1)\) varies between these best\hyp{}evolved individuals. 
We find the whole range from ground sensor~\(g(t)\) and previous opinion~\(o(t-1)\) having a strong impact to both sensors having only a low influence on the decision. 

In the second evolved decision\hyp{}making approach, the ground sensor~\(g(t)\) has the highest impact. 
We find two best\hyp{}evolved individuals where this is the case. 
In one of the two decision\hyp{}making mechanisms, the opinions of a robot's neighbors~\(w(t) + l(t)\) have a stronger impact than the robot's previous opinion~\(o(t-1)\). 
In the other decision\hyp{}making mechanism, the robot's previous opinion~\(o(t-1)\) and the opinions of a robot's neighbors~\(w(t) + l(t)\) have similar impact, see Fig.~\ref{fig:shap3}. 

As a third approach, we find one best\hyp{}evolved individual where the robot's previous opinion~\(o(t-1)\) has the highest impact, but ground sensor~\(g(t)\) and the opinions of a robot's neighbors~\(w(t) + l(t)\) have only a slightly lower influence (see Fig.~\ref{fig:shap2}). 

Overall, a high influence of the opinions of a robot's neighbors~\(w(t) + l(t)\) is to be favored by evolution, while the impact of the ground sensor~\(g(t)\) and the robot's previous opinion~\(o(t-1)\) varies. 
In Sec.~\ref{sec:efficiency}, we found that the best\hyp{}evolved individuals of the evolutionary runs 3 and 8 scaled best with problem difficulty. 
The decision\hyp{}making mechanism of run 3 follows the second strategy, i.e., the ground sensor~\(g(t)\) has the highest impact, see Fig.~\ref{fig:shap3}. 
While this strategy leads to high accuracy (i.e., high exit probability~\(E_N\)), it is slower than the best\hyp{}evolved individual of run 8. 
The decision\hyp{}making mechanism of run 8 follows the first strategy, where the opinions of a robot's neighbors~\(w(t) + l(t)\) have the strongest impact, see Fig.~\ref{fig:shap8}. 
In this case, the ground sensor~\(g(t)\) and the robot's previous opinion~\(o(t-1)\) have only a small influence on the decision. 
Based on these insights, we will give the opinions of a robot's neighbors~\(w(t) + l(t)\) the highest impact in our hand\hyp{}coded decision\hyp{}making mechanisms.

\subsubsection{Network Input Values}
\label{sec:inputs}

\begin{figure}[t]
    \centering
    \subfloat[dominant feature: \textit{Black}]{\includegraphics[width=0.5\linewidth]{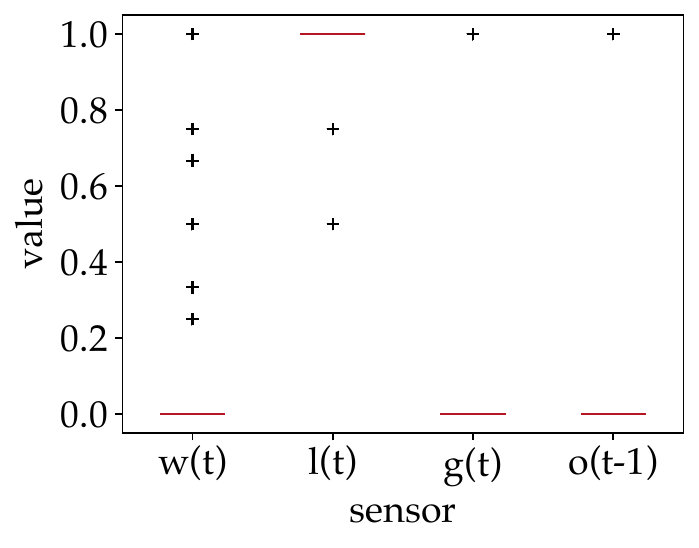}}
    \subfloat[dominant feature: \textit{White}]{\includegraphics[width=0.5\linewidth]{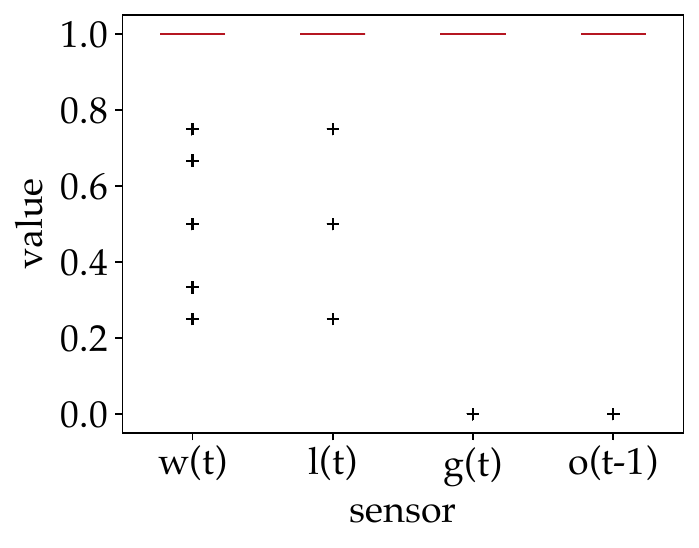}}
    \caption{Sensor values of all robots for one representative best\hyp{}evolved decision\hyp{}making mechanism with (a) \textit{Black} and (b) \textit{White} as the more frequent environmental feature. \(w(t)\)~gives the percentage of neighbors with opinion \textit{White}, \(l(t)\)~the normalized length of the message queue of neighbor opinions, \(g(t)\)~the current ground sensor value, and \(o(t-1)\) the robot's previous opinion.}
    \label{fig:inputs}
\end{figure}

In the next step, we analyze the distribution of the input values during an evaluation to investigate whether the input values vary or are rather static. 
Fig.~\ref{fig:inputs} visualizes the sensor values of all robots for a run of~\(200\)~s when using one of the evolved decision\hyp{}making mechanisms in problem difficulty~\(\rho^* = 0.25\). 
It is representative of all evolved decision\hyp{}making mechanisms and evaluation runs. 
We find that the percentage of neighbors with opinion \textit{White}~\(w(t)\) has a value corresponding to the more dominant environmental feature (i.e., \(0\) for \textit{Black} and \(1\) for \textit{White}). 
But the normalized length of the message queue~\(l(t)\) has a value of~\(1.0\) most of the time, regardless of the more dominant environmental feature. 
Consequently, \(l(t)\)~plays a similar role as a bias value in calculating the decision output. 
For this reason, we do not include the normalized length of the message queue~\(l(t)\) in our hand\hyp{}coded decision\hyp{}making mechanism and focus on the percentage of neighbors with opinion \textit{White}~\(w(t)\).
The values of ground sensor~\(g(t)\) and the robot's previous opinion~\(o(t-1)\) match mainly the dominant feature and have a similar value distribution. 
Since the ground sensor~\(g(t)\) often has a high impact in the evolved decision\hyp{}making mechanisms (see Sec.~\ref{sec:impact}), we will integrate it as a second factor in the hand\hyp{}coded decision\hyp{}making mechanisms and exclude the robot's previous opinion~\(o(t-1)\).

\subsection{New Decision\hyp{}Making Mechanisms} 
\label{sec:new_cdm}

\subsubsection{Implementation}
We hand\hyp{}code two new decision\hyp{}making mechanisms based on the insights gained from our analysis. 
As discussed above, we will integrate both the percentage of neighbors with opinion \textit{White}~\(w(t)\), and ground sensor value~\(g(t)\), but \(g(t)\) will have a much lower impact than~\(w(t)\) (see Sec.~\ref{sec:impact}). 
Since \(w(t)\) is normalized, the new decision-making mechanisms are independent of queue sizes. 
We propose two new decision\hyp{}making mechanisms: (i)~a weighted sum of the two sensor inputs and (ii)~a decision\hyp{}making mechanism based on conditional statements.  

We define decision\hyp{}making mechanism HC1 as 

\begin{equation}
    o(t) = \begin{cases}
        1 & \text{if } 0.75 w(t) + 0.25 g(t) >= 0.5 \\
        0 & \text{otherwise}
    \end{cases} \, , 
    \label{eq:HC1}
\end{equation}

with robot opinion~\(o(t)\), percentage of neighbors with opinion \textit{White}~\(w(t)\), and ground sensor value~\(g(t)\) at the current time step~\(t\). 
It is a weighted sum that weights the percentage of neighbors with opinion \textit{White}~\(w(t)\) three times higher than the ground sensor value~\(g(t)\). 
Thus, \(g(t)\) has a low impact, similar to the best\hyp{}evolved individual of run~8. 
Intuitively speaking, a robot will switch directly to opinion \textit{White} if at least two\hyp{}thirds of its neighbors have \textit{White} as their current opinion (i.e., \(w(t) \ge \frac{2}{3}\)). 
A robot will directly adopt opinion \textit{Black} if less than one\hyp{}third of its neighbors have opinion \textit{White} (i.e., \(w(t) < \frac{1}{3}\)). 
The robot's ground sensor value~\(g(t)\) will influence the decision output~\(o(t)\) only in the other cases, i.e., when there is no clear majority for one environmental feature. 
A drawback of this approach is that there is an imbalance in the adoption of \textit{White} or \textit{Black} as the current opinion.
This is because a robot will adopt \textit{White} as its opinion if two\hyp{}thirds of its neighbors have opinion \textit{White} (i.e., \(w(t) = \frac{2}{3}\)) regardless of the ground sensor value~\(g(t)\), but it will only switch to opinion \textit{Black} if two\hyp{}thirds of its neighbors have opinion \textit{Black} (i.e., \(w(t) = \frac{1}{3}\)) and the ground sensor value~\(g(t)\) is also \textit{Black}. 
This imbalance is caused by the encoding of the neighbor opinions as the percentage of neighbors with opinion~\textit{White}~\(w(t)\). 
We will investigate the effect of this imbalance on the decision\hyp{}making in our benchmarks in the next section.  

\begin{algorithm}[t]
    \caption{decision\hyp{}making mechanism HC2}\label{alg:HC2}
\begin{algorithmic}[1]
    \If{$l(t) == 0 $} \Comment{no neighbor opinions received}
        \State $o(t) \gets o(t-1)$ \Comment{keep own opinion}
    \Else 
    \If{$w(t) \ge 0.75$} \Comment{majority has opinion \textit{White}}
    \State $o(t) \gets 1$ \Comment{new opinion: \textit{White}}
    \ElsIf{$w(t) \le 0.25$} \Comment{majority has opinion \textit{Black}}
    \State $o(t) \gets 0$ \Comment{new opinion: \textit{Black}}
    \Else 
    \If{$g(t) == 1$} \Comment{ground sensor value: \textit{White}}
        \State $o(t) \gets 1$ \Comment{new opinion: \textit{White}}
    \Else \Comment{ground sensor value: \textit{Black}}
        \State $o(t) \gets 0$ \Comment{new opinion: \textit{Black}}
    \EndIf
\EndIf 
\EndIf
\end{algorithmic}
\end{algorithm}

To remove the inherent bias for opinion~\textit{White} in the weighted sum, we propose decision\hyp{}making mechanism HC2 that is based on conditional statements, see Alg.~\ref{alg:HC2}. 
HC2 works similarly to the weighted sum of HC1, except that the ground sensor~\(g(t)\) is the deciding factor when one\hyp{}third to two\hyp{}thirds of the robot's neighbors have opinion \textit{White} (i.e., \(\frac{1}{3} \le w(t) \le \frac{2}{3}\)).

\subsubsection{Efficiency}
\label{sec:Results_Benchmarks}

\begin{table*}[t]
	\caption{Mean consensus times~$\overline{T}_N$ and exit probabilities~$E_N$ for the benchmarks (see Sec.~\ref{sec:benchmarks}) with the voter model (VM), the majority rule (MR), the evolved decision\hyp{}making mechanisms (EVO), and our two hand\hyp{}coded collective decision\hyp{}making mechanisms HC1~(Eq.~\ref{eq:HC1}) and HC2~(Alg.~\ref{alg:HC2}). For EVO, the median values of the 10 best\hyp{}evolved individuals are given. 
\label{tab:metrics}
}
\centering
		\begin{tabular}{rrrrrrrrrrr}
		\hline 
		{dominant} & {decision\hyp{}making}
		& \multicolumn{2}{c}{$\rho^* =0.25$} & \multicolumn{2}{c}{$\rho^* =0.52$} & \multicolumn{2}{c}{$\rho^* =0.67$} & \multicolumn{2}{c}{$\rho^* =0.82$} \\
		feature & mechanism & & & & & & & & 
		\\ 
		 & & \multicolumn{1}{c}{$\overline{T}_N$} &  \multicolumn{1}{c}{$E_N$}&  \multicolumn{1}{c}{$\overline{T}_N$} &  \multicolumn{1}{c}{$E_N$} & \multicolumn{1}{c}{$\overline{T}_N$} &  \multicolumn{1}{c}{$E_N$} &  \multicolumn{1}{c}{$\overline{T}_N$} & \multicolumn{1}{c}{$E_N$}\\ 
	   \hline 
		\multirow{5}{*}{\begin{turn}{90}\textit{Black}\end{turn}} & VM & $94.9~\text{s}$ & $100$~\%  & $154.7~\text{s}$ & $97.1~\text{\%}$ & $192.5~\text{s}$ & $83.4~\text{\%}$  &  $206.3~\text{s}$ & $60.2~\text{\%}$ \\ 
		& MR & $69.5~\text{s}$ & $96.7~\text{\%}$ & $84.5~\text{s}$ & $83.5~\text{\%}$ & $86.8~\text{s}$ & $72.8~\text{\%}$  &  $94.1~\text{s}$ & $61.1~\text{\%}$ \\ 
        & EVO & $55.7~\text{s}$ & $100~\text{\%}$ & $85.1~\text{s}$ & $99.6~\text{\%}$ & $111.5~\text{s}$ & $98.0~\text{\%}$  &  $140.1~\text{s}$ & $85.8~\text{\%}$ \\
        & HC1 & $52.9~\text{s}$ & $100$~\% & $75.7~\text{s}$ & $98.1~\text{\%}$ & $92.8~\text{s}$ & $88.8~\text{\%}$ &$109.9~\text{s}$ & $69.5~\text{\%}$ \\ 
        & HC2 & $50.6~\text{s}$ & $100$~\% & $72.2~\text{s}$ & $99.0$~\% & $89.3~\text{s}$ & $92.9$~\% & $108.2$~s & $77.4$~\%\\ 
        \hline 
		\multirow{5}{*}{\begin{turn}{90}\textit{White}\end{turn}} & VM & $94.9~\text{s}$ & $100$~\% & $154.7~\text{s}$ & $97.1~\text{\%}$ & $192.5~\text{s}$ & $83.4~\text{\%}$  &  $206.3~\text{s}$ & $60.2~\text{\%}$ \\ 
		& MR & $69.5~\text{s}$ & $96.7~\text{\%}$ & $84.5~\text{s}$ & $83.5~\text{\%}$ & $86.8~\text{s}$ & $72.8~\text{\%}$  &  $94.1~\text{s}$ & $61.1~\text{\%}$ \\ 
        & EVO & \(54.8\)~s & $100~\text{\%}$ & \(87.2\)~s & $98.9~\text{\%}$ & \(115.3\)~s & $87.4~\text{\%}$ & \(129.6\)~s & $59.2~\text{\%}$ \\ 
        & HC1 & $51.5~\text{s}$ & $100$~\% & $70.0~\text{s}$ & $99.2$~\% & $87.0~\text{s}$ & $93.9$~\% & $103.1~\text{s}$ & $80.9$~\% \\ 
        & HC2  & $51.1~\text{s}$ & $100$~\% & $73.2~\text{s}$ & $99.3$~\% &  $90.8~\text{s}$ & $92.5$~\% & $109.4$~s & $75.0$~\%\\ 
		\hline 
	\end{tabular}
\end{table*}
\begin{figure*}
    \centering
     \subfloat[$E_N$ for $\rho^* = 0.25$\label{fig:ep025}]{\includegraphics[width=0.25\linewidth]{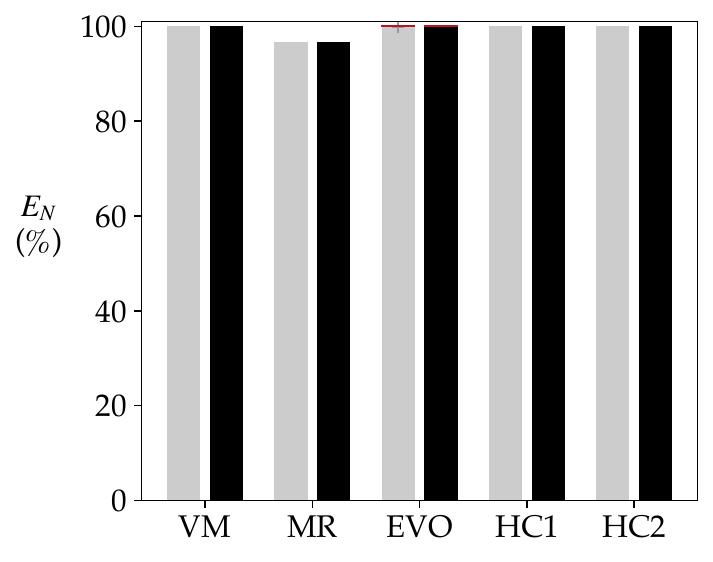}}
    \subfloat[$E_N$ for $\rho^* = 0.52$\label{fig:ep052}]{\includegraphics[width=0.25\linewidth]{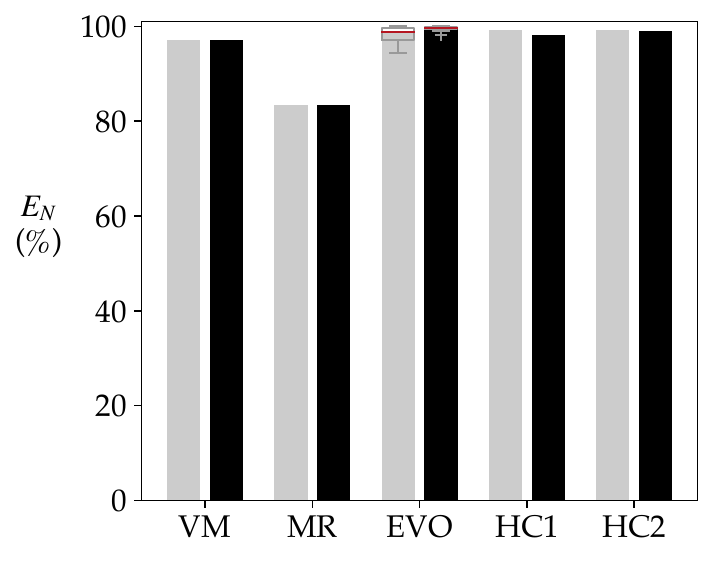}}
    \subfloat[$E_N$ for $\rho^* = 0.67$\label{fig:ep067}]{\includegraphics[width=0.25\linewidth]{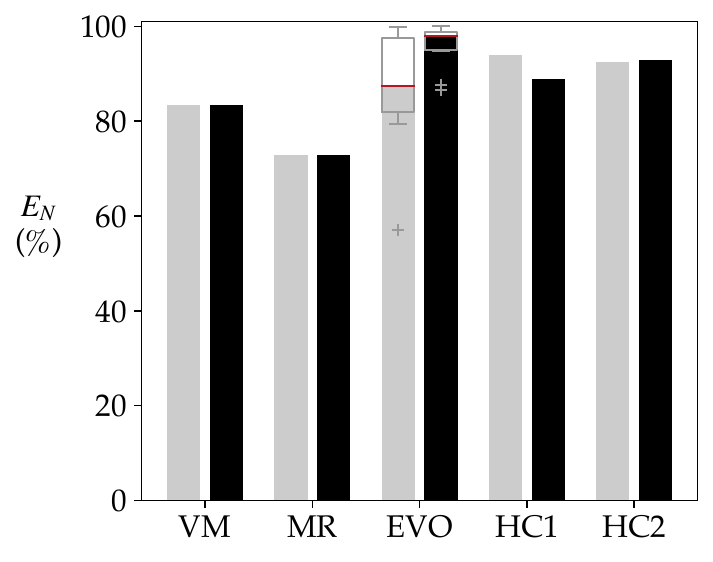}}
    \subfloat[$E_N$ for $\rho^* = 0.82$\label{fig:ep082}]{\includegraphics[width=0.25\linewidth]{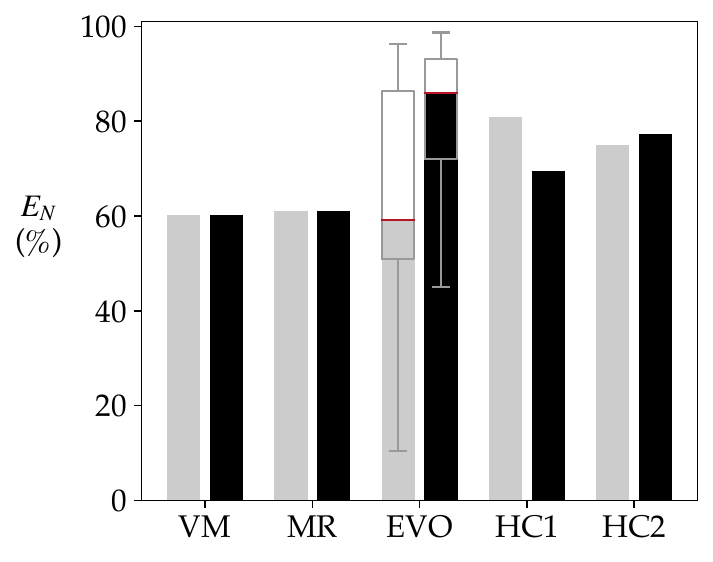}} \\
    \subfloat[$\overline{T}_N$ for $\rho^* = 0.25$\label{fig:ct025}] {\includegraphics[width=0.25\linewidth]{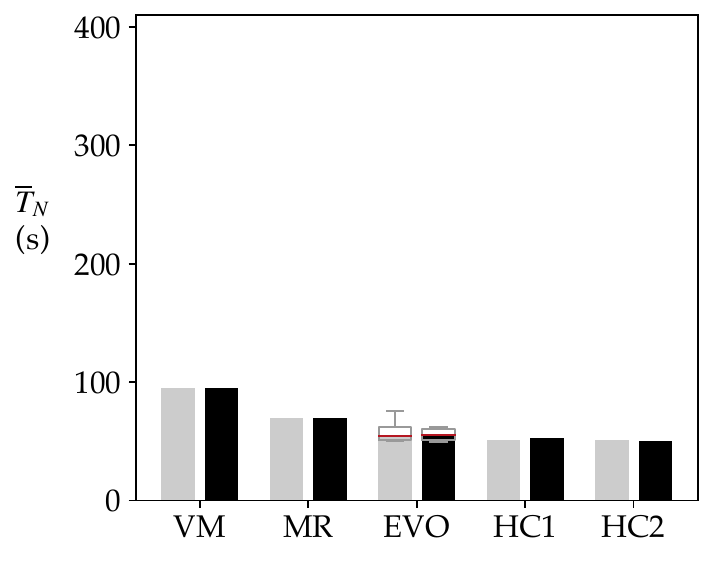}}
    \subfloat[$\overline{T}_N$ for $\rho^* = 0.52$\label{fig:ct052}]{\includegraphics[width=0.25\linewidth]{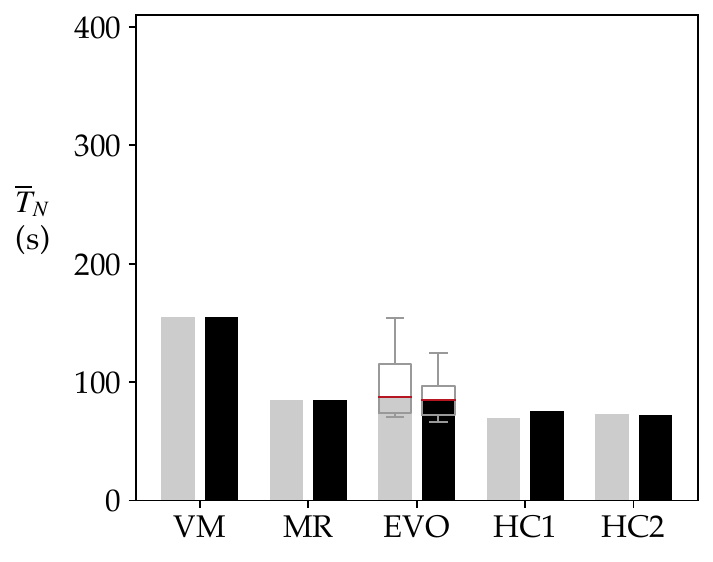}}
    \subfloat[$\overline{T}_N$ for $\rho^* = 0.67$\label{fig:ct067}]{\includegraphics[width=0.25\linewidth]{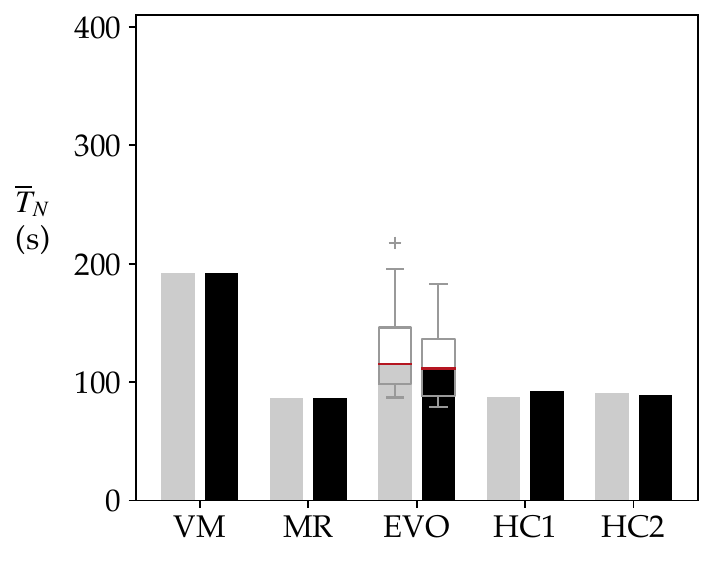}}
    \subfloat[$\overline{T}_N$ for $\rho^* = 0.82$\label{fig:ct082}]{\includegraphics[width=0.25\linewidth]{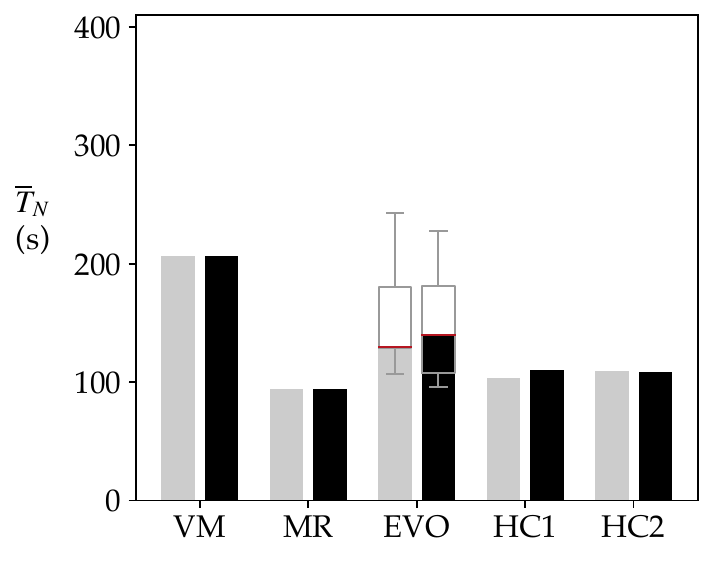}}
    \caption{Exit probabilities~$E_N$ (higher is better) and mean consensus times~$\overline{T}_N$ (lower is better) for the benchmark runs with the voter model (VM), the majority rule (MR), the evolved decision\hyp{}making mechanisms (EVO), and our two hand\hyp{}coded decision\hyp{}making mechanisms HC1~(Eq.~\ref{eq:HC1}) and HC2~(Alg.~\ref{alg:HC2}) for \textit{White}\hyp{}dominant (left) and \textit{Black}\hyp{}dominant (right) environments. For EVO the median value as well as a boxplot of the median values of the 10 best\hyp{}evolved individuals are included.}
    \label{fig:consensustime}
\end{figure*}

We compare the two new hand\hyp{}coded decision\hyp{}making mechanisms HC1 and HC2 with our evolved decision\hyp{}making mechanisms, the voter model, and the majority rule in our benchmark settings (see Sec.~\ref{sec:benchmarks}). 
Mean consensus times~\(\overline{T}_N\) and exit probabilities~\(E_N\) are given in Tab.~\ref{tab:metrics} and are visualized in Fig.~\ref{fig:consensustime} for the four different problem difficulties~\(\rho^* \in \{0.25, 0.52, 0.67, 0.82\}\). 
Here, we compare against the median performance of the evolved decision-making mechanisms as a representative value.
In Sec.~\ref{sec:efficiency}, we found differences in the efficiency of the evolved decision\hyp{}making mechanisms for \textit{White}\hyp{}dominant and \textit{Black}\hyp{}dominant environments with increasing problem difficulty. 
Similarly, the hand\hyp{}coded decision\hyp{}making mechanism HC1 does not perform equally well for \textit{White}\hyp{}dominant and \textit{Black}\hyp{}dominant environments (p\(< 0.05\)). 
This is intuitive, since the weighted sum implementing HC1 has an inherent preference for \textit{White}.
We find no significant performance differences for HC2 depending on the dominant environmental feature. 
Comparing HC1 and HC2, we find differences in the mean consensus time~\(\overline{T}_N\) of up to \(6\)~s.
HC1 is significantly faster in \textit{White}\hyp{}dominant environments for problem difficulties~\(\rho^* \in \{0.52, 0.67, 0.82\}\) (p \(< 0.05\)) while HC2 leads to higher decision accuracy in \textit{Black}\hyp{}dominant environments. 
Compared to the evolved decision\hyp{}making mechanisms, HC1 and HC2 have lower exit probabilities~\(E_N\) for \textit{Black}\hyp{}dominant environments and higher exit probabilities for \textit{White}\hyp{}dominant environments. 
The difference increases with increasing problem difficulty~\(\rho^*\). 
HC1 and HC2 are also up to \(32\)~s faster than the evolved decision\hyp{}making mechanisms for both \textit{White}\hyp{}dominant and \textit{Black}\hyp{}dominant environments. 
Both hand\hyp{}coded decision\hyp{}making mechanisms outperform the voter model for all problem difficulties in decision speed (i.e., lower mean consensus times~\(\overline{T}_N\)). 
We find that HC1 and HC2 are faster than the majority rule for the easier problem difficulties~\(\rho^* \in \{0.25, 0.52\}\), but equally fast or slower for the harder problem difficulties~\(\rho^* \in \{0.67, 0.82\}\) (p\(<0.05\)). 
The exit probabilities~\(E_N\) of HC1 and HC2 are higher than for the voter model and the majority rule for all problem difficulties, i.e., the hand\hyp{}coded decision\hyp{}making mechanisms are more accurate than the state\hyp{}of\hyp{}the\hyp{}art collective decision\hyp{}making mechanisms. 
Thus, we find that the two new hand\hyp{}coded decision\hyp{}making mechanisms are competitive with or even more efficient than the voter model and the majority rule.
Overall, the two new hand\hyp{}coded decision\hyp{}making mechanisms bring advantages in terms of efficiency compared to the voter model and the majority rule, and have higher interpretability than the evolved decision\hyp{}making mechanisms while having lower exit probabilities~\(E_N\) and shorter mean consensus times~\(\overline{T}_N\).

\section{Conclusion}

Evolved decision\hyp{}making behaviors usually have lower interpretability than hand\hyp{}coded decision\hyp{}making behaviors but offer more efficiency in terms of decision speed and accuracy. 
In this paper, we have shown that applying methods for explainable AI can help us determine how the evolved decision\hyp{}making mechanisms work. 
Based on the insights gained, we can implement new decision\hyp{}making mechanisms that offer higher interpretability than the evolved artificial neural networks and are more efficient than the state\hyp{}of\hyp{}the\hyp{}art mechanisms voter model and majority rule. 

Although we gained enough insight into the evolved decision\hyp{}making mechanisms to hand\hyp{}code improved decision\hyp{}making mechanisms, the analysis did not reveal the complete strategy of the evolved decision\hyp{}making mechanisms. 
Here, our goal was not to gain a complete understanding of the evolved decision\hyp{}making mechanisms, but to gain enough insight to implement better decision\hyp{}making mechanisms. 
Analyzing the evolved decision\hyp{}making mechanisms in more detail to determine the exact way the mechanisms work is an interesting aspect for future research. 

We still find the speed versus accuracy trade\hyp{}off in both  the evolved decision\hyp{}making mechanisms and the hand\hyp{}coded decision\hyp{}making mechanisms. 
However, it is less pronounced than for the voter model and the majority rule. 
There are potentially more simple and more efficient decision\hyp{}making mechanisms that do not rely on more communication between robots but additionally only take into account the robots' own sensor measurements. 
Works on decision\hyp{}making for estimation tasks, i.e., without a desired unique consensus as in the best\hyp{}of\hyp{}\(n\) problem, already take into account the robots' individual sensor measurements~\cite{10161354} and our study shows that it can also be beneficial for cases where a global consensus is desired. 

The voter model and the majority rule can be easily scaled from best\hyp{}of\hyp{}\(2\) problems to best\hyp{}of\hyp{}\(n\) problems. 
A drawback of the evolved and the newly hand\hyp{}coded decision\hyp{}making mechanisms is that they cannot be applied as easily to best\hyp{}of\hyp{}\(n\) problems. 
The evolved decision\hyp{}making mechanisms are most likely not applicable at all to scenarios with a different \(n\), and new specialized decision\hyp{}making mechanisms must be evolved. 
The new hand\hyp{}coded algorithms need some adaptation, with the conditional statement\hyp{}based HC2 possibly being easier to modify than the weighted sum of HC1. 

Further studies that we leave for future work are evolving the weights for the neighbor opinions and the ground sensor value, varying the size of the multi-robot system, and testing the mechanisms in the benchmark settings proposed by~Bartashevich and Mostaghim~\cite{bartashevich2019}.

\printbibliography

\end{document}